\documentstyle[leqno]{bcp}

 1   
\font\ddpp=msbm9  scaled \magstep 1  
 1  

\def\R{\hbox{\ddpp R}}               

\def\hfl#1#2{\smash{\mathop{\hbox to 12 mm{\rightarrowfill}}
\limits^{\scriptstyle#1}_{\scriptstyle#2}}}

\begin{document}


\mathclass{Primary 70S05; Secondary 58A20.}

\thanks{This work has been partially supported by grants BFM2001-2272 and
BFM2000-0808. }

\abbrevauthors{M. de Le{\'o}n, J.C. Marrero and D. Mart{\'i}n de Diego}

\abbrevtitle{A new geometric setting for classical
field theories}

\title{A new geometric setting\\ for classical
field theories}

\author{M.\ de\ Le{\'o}n}

\address{Instituto de Matem\'aticas y F\'\i sica Fundamental\\
Consejo Superior de Investigaciones Cient\'\i ficas\\
Serrano 123, 28006 Madrid, Spain\\
E-mail: mdeleon@imaff.cfmac.csic.es\\
}

\author{J.\ C.\ Marrero}

\address{Departamento de Matem\'atica Fundamental\\
Facultad de Matem\'aticas,
Universidad de La Laguna\\
La Laguna, Tenerife, Canary Islands, Spain\\
E-mail: jcmarrer@ull.es}

\author{D.\ Mart{\'i}n\ de\ Diego}

\address{Instituto de Matem\'aticas y F\'\i sica Fundamental\\
Consejo Superior de Investigaciones Cient\'\i ficas\\
Serrano 123, 28006 Madrid, Spain\\
E-mail: d.martin@imaff.cfmac.csic.es}

\maketitlebcp

\abstract{A new geometrical setting for classical field theories
is introduced. This description is strongly inspired in the one due to
Skinner and Rusk for singular lagrangians systems. For a singular field theory a constraint
algorithm is developed that gives a final constraint submanifold where a well-defined dynamics
exists. The main advantage of this algorithm is that
the second order condition is automatically included.
}

\section*{1. Introduction.}
The search of a convenient setting for classical field theories
has been an strong motivation for geometers and physicists in the last forty years.
In the end of the sixties it was developed the so-called multisymplectic
formalism, which is a natural extension of the symplectic
framework for mechanics.

The multisymplectic approach was developed by the Polish school
led by W. Tulczyjew (see \cite{BSF} for more details), and
independently by P.L Garc{\'\i}a and A. P\'erez-Rend\'on
\cite{sala1,sala2}, and Goldschmidt and Sternberg \cite{GoSt}.
This approach leads to a geometric definition of multisymplectic
form in \cite{gimmsy1,gimmsy2}, and more recently in
\cite{CIL2,CIL1} where a careful study of these structures is
developed (see also \cite{martin1,martin2} for previous results,
and \cite{torun,aitor,PR1,PR2} for recent developments).

There are two different ways to present the evolution equations in a
geometric form. One uses the notion of Ehresmann connections
\cite{LMMS,LMMS2} which is widely employed along the present paper.
The other one uses the notion of multivector field
(see \cite{barsa01,barsa02,barsa03,barsa04}).
Of course, both are equivalent, and permit to develop
a convenient constraint algorithm when we are dealing
with singular lagrangians.

Alternative geometric approaches based on the so-called $n$-symplectic
geometry (see \cite{survey} for a recent survey), and
polysymplectic geometry (see \cite{gena1,gena2}) are also available.

The aim of the present paper is to give a new geometric setting,
based in that developed by Skinner and Rusk \cite{SR1,SR2}.
In order to treat with singular lagrangian systems, Skinner and Rusk
have constructed a hamiltonian system on the Withney sum
$T^*Q \oplus TQ$ over the configuration manifold $Q$.
The advantage of their approach lies on the fact that
the second order condition of the dynamics is automatically satisfied.
This does not happen in the Gotay and Nester formulation, where the second
order condition problem has to be considered after the
implementation of the constraint algorithm (see \cite{GN1,GN2,LM}).

Here, we start with a lagrangian function defined on $Z$,
where $\pi_{XZ} : Z \longrightarrow X$ is the 1-jet prolongation
of a fibration $\pi_{XY} : Y \longrightarrow X$.
We consider the fibration $\pi_{XW_0} : W_0 \longrightarrow X$,
where $W_0 = \Lambda^n_2Y \times_Y Z$ is the fibered product.
On $W_0$ we construct a multisymplectic form by pulling back the
canonical multisymplectic form on $\Lambda^n_2Y$, and define a convenient
hamiltonian. The solutions of the field equations
are viewed as integral sections of Ehresmann connections in the fibration
$\pi_{XZ} : Z \longrightarrow X$.
The resultant algorithm is compared with the ones developed
in the lagrangian and hamiltonian sides.
The scheme is applied to an example, the bosonic string.
The case of time-dependent mechanics is recovered as a particular case.
The paper also contains three appendices exhibiting some notions
and properties of Ehresmann connections.

\section*{2. Lagrangian formalism.}
A classical field theory consists of a fibration $\pi_{XY} : Y \longrightarrow X$
(that is, $\pi_{XY}$ is a surjective submersion)
over an orientable $n$-dimensional manifold $X$ and an
$n$-form $\Lambda$ (the lagrangian form) defined on the 1-jet prolongation
$\pi_{XZ} : J^1\pi_{XY} \longrightarrow X$ along
the projection $\pi_{XY}$. We will use the notation $Z = J^1 \pi_{XY}$.
In addition, if $\eta$ is a fixed volume form on $X$ we have
$\Lambda = L \eta$, where $L$ is a function on $Z$.
An additional fiber bundle $\pi_{YZ} : Z \longrightarrow Y$ is also obtained.
Here $X$ represents the space-time manifold,
and the fields are viewed as sections of $\pi_{XY}$. (See \cite{BSF,gimmsy1,gimmsy2,KT,gena1,gena2}).

\defin{Definition}{2.1.
A lagrangian $L:Z \longrightarrow \R$ is said to be regular if
the hessian matrix
$$
\left(\frac{\partial^2 L}{\partial z^i_\mu \partial z^j_\nu}\right)
$$
is regular. Otherwise, $L$ is said to be singular.}

Along this paper we will choose fibered coordinates
$(x^\mu, y^i, z^i_\mu)$ on $Z$ such that
$\eta = d^nx = dx^1 \wedge \dots \wedge dx^n$. Here $\mu$ runs from
$1$ to $n$, and $i$ runs from $1$ to $m$, so that
$Y$ has dimension $n+m$. A useful notation will be
$d^{n-1}x^{\mu}= i_{\frac{\partial}{\partial x^\mu}}\eta$.

The volume form $\eta$ permits to construct a tensor field of type $(1,n)$ on $Z$:
$$
S_{\eta} = (dy^i - z^i_{\mu} dx^{\mu}) \wedge d^{n-1}x^{\nu} \otimes
\frac{\partial}{\partial z^i_{\nu}}.
$$

Next, the {\sl Poincar\'e-Cartan $n$-form and $(n+1)$-form} are defined as follows:
$$
\Theta_L = \Lambda + S_{\eta}^* (dL) \; , \quad
\Omega_L = - d \Theta_L,
$$
where $S_{\eta}^*$ is the adjoint operator of $S_{\eta}$.
In coordinates, we have
\begin{eqnarray*}
\Theta_L & = & (L-z^i_\mu \frac{\partial L}{\partial z^i_\mu}) d^nx
+ \frac{\partial L}{\partial z^i_\mu} dy^i \wedge d^{n-1}x^{\mu}\\
\Omega_L & = & - d(L-z^i_\mu \frac{\partial L}{\partial z^i_\mu}) \wedge
d^nx - d(\frac{\partial L}{\partial z^i_\mu}) \wedge dy^i \wedge
d^{n-1}x^{\mu}.
\end{eqnarray*}

An {\sl extremal} of $L$ is a section $\phi$ of $\pi_{XY}$ such that, for any vector $\xi_Z$ on $Z$,
\begin{equation}\label{fieldequation}
(j^1\phi)^*(i_{\xi_Z}\Omega_L)=0
\end{equation}
where $j^1\phi$ is the first jet prolongation of $\phi$.

As is well-known, $\phi$ is an extremal of $L$ if and only if it satisfies
the Euler-Lagrange equations:
\begin{equation}\label{asd}
(j^1\phi)^*\left( \frac{\partial L}{\partial
y^i}-\frac{d}{dx^{\mu}}\left(\frac{\partial L}{\partial
z^i_{\mu}}\right)\right)=0, \qquad 1\leq i\leq n.
\end{equation}

We can consider a more general kind of solutions, those sections
$\psi$ of the fiber bundle $\pi_{XZ} : Z \longrightarrow X$
such that
\begin{equation}\label{deDonderequation}
\psi^*(i_{\xi_Z}\Omega_L)=0 \; ,
\end{equation}
for any vector $\xi_Z$ on $Z$.
Equation (\ref{deDonderequation}) is referred as the de Donder equations.

Looking at (\ref{deDonderequation}) we have an alternative characterization.
Let $\Gamma$ be an Ehresmann connection in $\pi_{XZ}: Z\longrightarrow X$,
with horizontal projector ${\bf h}$. Consider the equation
\begin{equation}\label{connection1}
i_{\bf h}\Omega_L=(n-1)\Omega_L .
\end{equation}
The horizontal sections (if they exist) of $\Gamma$ are just the solutions of
the de Donder problem.

Indeed, if
$$
{\bf h}(\frac{\partial}{\partial x^\mu}) = \frac{\partial}{\partial x^\mu}
+ \Gamma^i_{\mu} \frac{\partial}{\partial y^i} +
\Gamma^i_{\nu \mu} \frac{\partial}{\partial z^i_\nu}
$$
then a direct computation shows that equation (\ref{connection1}) holds
if and only if
\begin{eqnarray}\label{uno}
(\Gamma^j_{\nu} - z^j_{\nu})
\left(\frac{\partial^2 L}{\partial z^i_\mu \partial z^j_\nu}\right) = 0 \\
\label{dos}
\frac{\partial L}{\partial y^i} - \frac{\partial^2 L}{\partial x^\mu \partial z^i_\mu}
- \Gamma^j_\mu \frac{\partial^2L}{\partial y^j \partial z^i_\mu}
- \Gamma^j_{\mu \nu} \frac{\partial^2 L}{\partial z^j_\nu \partial z^i_\mu} +
(\Gamma^j_{\nu} - z^j_{\nu}) \frac{\partial^2 L}{\partial y^i \partial z^j_\nu} = 0
\end{eqnarray}
(see \cite{LMMS}).

If the lagrangian $L$ is regular, then Eq. (\ref{uno}) implies that
$\Gamma^i_\mu = z^i_\mu$ and therefore (\ref{dos}) becomes
\begin{equation}\label{tres}
\frac{\partial L}{\partial y^i} - \frac{\partial^2 L}{\partial x^\mu \partial z^i_\mu}
- z^j_\mu \frac{\partial^2L}{\partial y^j \partial z^i_\mu}
- \Gamma^j_{\mu \nu} \frac{\partial^2 L}{\partial z^j_\nu \partial z^i_\mu}
= 0.
\end{equation}
Now, if $\tau(x^\mu)=(x^\mu, \tau^i(x), \tau^i_\mu(x))$ is an
integral section of $\Gamma$ we would have
$$
z^i_\mu=\frac{\partial \tau^i}{\partial x^\mu} \; \quad
\Gamma^i_{\mu \nu}=\frac{\partial \tau^i_\mu}{\partial x^\nu}
$$
which proves that Eq. (\ref{tres}) is nothing but the Euler-Lagrange equations
for $L$.

If the lagrangian $L$ is regular, then every solution $\psi$ of the de Donder equations
(\ref{deDonderequation}) is automatically a 1-jet prolongation, say
$\psi = j^1\phi$ and the section $\phi$ of
$\pi_{XY}$ is a solution of equations (\ref{fieldequation}).

In terms of Ehresmann connections, if $L$ is regular, then
any solution $\Gamma$ of equations (\ref{connection1}) is
semi-holonomic (see Appendix B).

\section*{3. Hamiltonian formulation.}
Let $\Lambda^n_r Y$, $1\leq r \leq m$, be the subbundle of the bundle $\Lambda^n Y$
of $n$-forms on $Y$ consisting of those $n$-forms which
vanish when $r$ of their arguments are vertical.
We have a chain of vector bundles over $Y$:
$$
0 \subset \Lambda^n_1 Y \subset \Lambda^n_2 Y \subset \cdots \subset \Lambda^nY
$$

The elements of $\Lambda_1^n Y$ (resp. $\Lambda_2^n Y$)
are locally expressed as
$p(x,y)d^nx$ (resp. $pd^n x+p^{\mu}_i dy^i\wedge d^{n-1} x^{\mu}$).
Thus, we introduce local coordinates $(x^{\mu}, y^i, p)$ on the manifold
$\Lambda_1^n Y$, and $(x^{\mu}, y^i, p, p^{\mu}_i)$ on $\Lambda_2^n Y$.

The manifold $\Lambda^nY$ carries a canonical $n$-form, $\Theta_0$,
which is defined as follows:
$$
\Theta_0 (\omega) (\xi_1,\xi_2, \ldots , \xi_n) =
\omega(\nu(\omega))(\nu_*(\xi_1), \nu_*(\xi_2),\ldots , \nu_*(\xi_n))
$$
where $\omega \in \Lambda^n Y$, $\xi_i \in T_{\omega}(\Lambda^nY)$,
and $\nu: \Lambda^n Y \longrightarrow Y$ is the canonical projection.

This form $\Theta_0$ induces an $n$-form on $\Theta_r$ on
$\Lambda_r^nY$, for each $r$, $1 \leq r \leq m$.

The closed $(n+1)$-forms $\Omega_r=-d\Theta_r$ (and of course,
$\Omega_0=-d\Theta_0$) are examples of
the so-called multisymplectic forms according the following definition.

\defin{Definition}{3.1.
A multisymplectic form on a manifold $M$ is a closed $k$-form $\Omega$
on $M$ such that the linear mapping $v \in T_xM \longrightarrow
i_v \Omega \in \Lambda^{k-1}T^*_xM$
is injective for all $x \in M$. The manifold $M$ equipped with a
multisymplectic form $\Omega$ will be called a multisymplectic
manifold, usually denoted by the pair $(M, \Omega)$.
Two multisymplectic manifolds $(M, \Omega)$ and
$(\bar{M}, \bar{\Omega})$ will be said multisymplectomorphic
if there exists a diffeomorphism $\phi : M \longrightarrow \bar{M}$
preserving the multisymplectic forms, say $\phi^* \bar{\Omega} = \Omega$;
$\phi$ will be called a multisymplectomorphism.
}

\remar{Remark}{3.2.
It will be useful to write the local expressions of the canonical
multisymplectic forms on $\Lambda_2^n Y$:
$$
\Theta_2 = p d^n x + p^\mu_i dy^i \wedge d^{n-1}x^{\mu},\quad
\Omega_2 = -dp \wedge d^n x - dp^\mu_i \wedge dy^i \wedge d^{n-1}x^{\mu}.
$$
}

A direct computation shows the following.

\th{Proposition}{3.3. \label{regularity}
Assume that $n \geq 2$. Then, a lagrangian $L$ is regular if and only
if the pair $(Z, \Omega_L)$ is a multisymplectic manifold.
}

Since $\Lambda_1^n Y$ is a vector subbundle of $\Lambda_2^n Y$
we can construct the quotient vector bundle
$\Lambda_2^n Y/\Lambda_1^n Y$ which will we denoted by $Z^*$.
The projection $\Lambda^n_2Y \longrightarrow Z^*$ will we denoted by $\lambda$.
We also have a fibration $\pi_{XZ^*} : Z^* \longrightarrow X$.

In this context, a hamiltonian $h$ is a section of $\lambda$.
Using this hamiltonian we define an $n$-form $\Theta_h$ on $Z^*$
by pulling back the canonical $n$-form $\Theta_2$, i.e.
$\Theta_h=h^*\Theta_2$. We put $\Omega_h=-d\Theta_h$ so that
$\Omega_h=h^*\Omega_2$.

A section $\sigma$ of $\pi_{XZ^*} : Z^* \longrightarrow X$
is said to satisfy the {\sl Hamilton equations} for a given hamiltonian $h$ if
\begin{equation}\label{fieldequationh}
\sigma^*(i_{\xi_{Z^*}}\Omega_h)=0 \, ,
\end{equation}
for all vector fields $\xi_{Z^*}$ on $Z^*$.

In local coordinates $(x^{\mu}, y^i, p^{\mu}_i)$ for $Z^*$, the section
$h$ may be represented by a local function $H$:
$$
p=-H(x^{\mu}, y^i, p^{\mu}_i)
$$
then
\begin{equation}\label{8'}
\Theta_h=-H d^n x + p^{\mu}_i dy^i\wedge d^{n-1}x^{\mu}, \quad
\Omega_h = dH \wedge d^n x - dp^{\mu}_i\wedge dy^i\wedge d^{n-1}x^{\mu},
\end{equation}
and the Hamilton equations for a section $\sigma$ become:
\begin{equation}\label{asd1}
\frac{\partial y^i}{\partial x^{\mu}}=\frac{\partial H}{\partial p^{\mu}_i}\; ,\qquad
\frac{\partial p^{\mu}_i}{\partial x^{\mu}}=-\frac{\partial H}{\partial y^i} \;.
\end{equation}

As in the precedent section, we can consider a connection $\widetilde\Gamma$
in $\pi_{XZ^*}: Z^* \longrightarrow X$, with horizontal projector $\widetilde{\bf h}$.
An intrinsic version of equations (\ref{asd1}) is then the following:
\begin{equation}\label{connection2}
i_{\bf \widetilde h}\Omega_h=(n-1)\Omega_h.
\end{equation}
Indeed, if $\widetilde{\Gamma}$ is flat, then its integral sections
are solutions of the Hamilton equations.

\remar{Remark }{3.4. If $n \geq 2$ then, from (\ref{8'}), it follows
that $\Omega_{h}$ is a multisymplectic form on $Z^{*}$.}

\section*{4. The Legendre transformation.}
Let $L$ be a lagrangian function. We define a fiber preserving map
$$
leg_L : Z \longrightarrow \Lambda^n_2 Y
$$
as follows:
$$
leg_L (j^1_x \phi)(X_1, \ldots, X_n)
= (\Theta_L)_{j_x^1\phi}(\tilde{X}_1, \ldots, \tilde{X}_n)
$$
for all $j^1_x\phi \in Z$ and $X_i \in T_{\phi(x)}Y$,
where $\tilde{X}_i \in T_{j^1_x\phi}Z$ are such that
$(\pi_{YZ})_*(\tilde{X}_i)=X_i$.

In local coordinates, we have
$$
leg_L(x^{\mu}, y^i, z^i_{\mu}) =
(x^{\mu}, y^i, p=L-z^{i}_{\mu}\frac{\partial L}{\partial z^{i}_{\mu}},
p^{\mu}_i=\frac{\partial L}{\partial z^{i}_{\mu}}).
$$

The Legendre transformation $Leg_L:Z \longrightarrow Z^*$ is defined as
the composition $Leg_L = \lambda \circ leg_L$, and it is locally expressed as
\begin{equation}\label{legendre}
Leg_L (x^\mu, y^i, z^i_\mu) = (x^\mu, y^i, \frac{\partial L}{\partial z^{i}_{\mu}}).
\end{equation}

>From the definitions, we deduce that $(leg_L)^*\Theta_2=\Theta_L$
and $(Leg_L)^*\Omega_2=\Omega_L$.

\th{Proposition}{4.1.
The lagrangian $L$ is regular if and only if the Legendre
transformation $Leg_L : Z \longrightarrow Z^*$ is a local diffeomorphism.
}

The Legendre transformation permits to connect the lagrangian and
hamiltonian descriptions as follows.

Assume the lagrangian $L$ be hyper-regular, that is, $Leg_ L : Z \longrightarrow
Z^*$ is a global diffeomorphism. We define a hamiltonian section
$h : Z^* \longrightarrow \Lambda^n_2Y$ by setting
$$
h = leg_L \circ (Leg_L)^{-1}.
$$
Then, from (\ref{legendre}) it follows that
$$
Leg_L^* \Theta_h = \Theta_L,\quad Leg_L^* \Omega_h = \Omega_L.
$$
Therefore, the solutions of equations (\ref{deDonderequation}) and (\ref{fieldequationh})
are $Leg_L$-related. In terms of connections, the solutions
of equations (\ref{connection1}) and
(\ref{connection2}) are also $Leg_L$-related.

If the lagrangian is regular, the equivalence is only at local level.
More precisely, if $n \geq 2$, we have that
$Leg_L$ is a local multisymplectomorphism between the
multisymplectic manifolds $(Z, \Omega_L)$ and $(Z^*, \Omega_h)$.

For singular lagrangians, a constraint algorithm was developed in
\cite{LMMS} (see Section 6).

\section*{5. A new geometric setting.}
Consider the fibered product $W_0 = \Lambda^n_2 Y \times_Y Z$
with canonical projections
$\hbox{pr}_1: W_0 \longrightarrow \Lambda^n_2 Y$ and
$\hbox{pr}_2: W_0 \longrightarrow Z$.
We consider fibered coordinates $(x^{\mu}, y^i, p, p^{\mu}_i, z_{\mu}^i)$
on $W_0$.

Define the $n$-form
$\Theta = \hbox{pr}_1^*\Theta_2$ and the $(n+1)$-form
$\Omega = -d\Theta = \hbox{pr}_1^*\Omega_2$.

We also define a function
$\Phi : W_0 \longrightarrow \R$ as follows.
Take an element $(\omega_{\phi(x)}, j_x^1\phi) \in W_0$,
then
$\Phi( (\omega_{\phi(x)}, j_x^1\phi))=a(x)$, where
$$
\phi^*(\omega_{\phi(x)}) = a(x) \eta(x).
$$
Locally, we have
$$
\Phi(x^{\mu}, y^i, p, p^{\mu}_i, z_{\mu}^i) = p+p^{\mu}_i z^i_{\mu}.
$$

Define also the function $H_0 : W_0 \longrightarrow \R$ by setting
$$
H_{0} = \Phi - \hbox{pr}_2^*L.
$$
The function $H_0$ locally reads as
$$
H_0(x^{\mu}, y^i, p, p^{\mu}_i, z_{\mu}^i) = p+p^{\mu}_i z^i_{\mu}
-L(x^{\mu}, y^i, z^i_{\mu}).
$$
Put
$$
\Omega_{H_0} = \Omega + dH_0 \wedge \eta.
$$
In local coordinates we have
$$
\Omega_{H_0} = -dp \wedge d^n x - dp^{\mu}_i \wedge dy^i\wedge d^{n-1}x^{\mu}
+ dH_{0}\wedge d^nx.
$$
Let $\bar{\Gamma}$ be an Ehresmann connection in the
fibered bundle $\pi_{X W_0}: W_0 \longrightarrow X$,
with horizontal projector $\bar{\hbox{\bf h}}$.

We search for a solution of the equation:
\begin{equation}\label{connection}
i_{\bar{\hbox{\bf h}}} \Omega_{H_0} = (n-1)\Omega_{H_0} \; .
\end{equation}
Define
\begin{eqnarray*}
W_1 & = & \{u \in W_0 \; /\; \exists  \bar{\hbox{\bf h}}_{u} : T_u W_0
\longrightarrow T_u W_0 \quad \hbox{linear such that }\
\bar{\hbox{\bf h}}^2_u = \bar{\hbox{\bf h}}_u,\\
&&\ker \bar{\hbox{\bf h}}_u = (V\pi_{X W_0})_u, \ i_{\bar{\hbox{\bf h}}_u}
\Omega_{H_0}(u) = (n-1) \Omega_{H_0}(u)\}.
\end{eqnarray*}

Suppose that the local expression of $\bar{\hbox{\bf h}}$
is
\begin{eqnarray*}
\bar{\hbox{\bf h}}(\frac{\partial}{\partial x^{\mu}})&
=&\frac{\partial}{\partial x^{\mu}}+
A^{i}_{\mu}\frac{\partial}{\partial y^i}+
B_{\mu}\frac{\partial}{\partial p}+C^{\nu}_{\mu i}\frac{\partial}{\partial p^{\nu}_i}+
D^{i}_{\mu \nu}\frac{\partial}{\partial z_{\nu}^i}\\
\bar{\hbox{\bf h}}(\frac{\partial}{\partial y^{i}})&=&0,\qquad
\bar{\hbox{\bf h}}(\frac{\partial}{\partial p})=0\\
\bar{\hbox{\bf h}}(\frac{\partial}{\partial p^{\mu}_i})&=&0,\qquad
\bar{\hbox{\bf h}}(\frac{\partial}{\partial z_{\mu}^i})=0
\end{eqnarray*}

We then obtain 
\begin{eqnarray*}
i_{\bar{\hbox{\bf h}}}\Omega_{H_0}& = & i_{\bar{\hbox{\bf h}}}
\left(-dp \wedge d^nx-dp^{\mu}_i \wedge dy^i \wedge d^{n-1}x^{\mu}
+ dH_0  \wedge d^nx \right)\\
&=& (n-1) \Omega_{H_0} + \left(C^{\mu}_{\mu i}-
\frac{\partial L}{\partial y^i} \right) dy^i \wedge d^nx\\
&&+
\left(z^i_{\mu}-A^i_{\mu}\right)\, dp_i^{\mu}\wedge d^nx
+ \left(p_i^{\mu}-\frac{\partial L}{\partial z^i_{\mu}}\right)\, dz^i_{\mu}\wedge d^n x
\end{eqnarray*}

Therefore, the submanifold $W_1$ of $W_0$
is determined by the vanishing of the constraints:
$$
p_i^{\mu}-\frac{\partial L}{\partial z^i_{\mu}}=0,
$$
and the components of the connection $\bar{\hbox{\bf h}}$ would verify
the following relations:
\begin{eqnarray}
A^i_{\mu}&=& z^i_{\mu} \label{unomas}\\
C^{\mu}_{\mu i}&=&\frac{\partial L}{\partial y^i} \label{dosmas}
\end{eqnarray}

>From the definition of $W_1$ we know that for each point $u \in
W_1$ there exists a ``horizontal projector'' $\bar{\hbox{\bf h}}_u
: T_uW_0 \longrightarrow T_u W_0$ satisfying equation
(\ref{connection}). However, we can not ensure that such $\bar{\bf
h}_u$, for each $u \in W_1$ will take values in $T_uW_1$.

But notice that the condition
$\bar{\hbox{\bf h}}_u(T_uW_0) \subset T_uW_1$, $\forall u \in W_1$
is equivalent to have
$$
\bar{\hbox{\bf h}}(\frac{\partial}{\partial x^{\mu}})\left(p^\kappa_j -
\frac{\partial L}{\partial z^j_\kappa}\right) = 0
$$
or, equivalently,
\begin{equation}\label{matrices1}
C^{\kappa}_{\mu j} = \frac{\partial^2 L}{\partial z^j_{\kappa} \partial x^{\mu}}
+ z_{\mu}^i \frac{\partial^2 L}{\partial z^j_{\kappa} \partial y^i}
+ D^{i}_{\mu\nu}\frac{\partial^2 L}{\partial z^j_{\kappa} \partial z^{i}_{\nu}}.
\end{equation}

We remark that if the lagrangian $L$ is regular, then
equations (\ref{matrices1}) have solutions $D$'s for a particular choice
of $C$'s satisfying equations (\ref{dosmas}).
Of course, we can take arbitrary values for the $B$'s.
A global solution is obtained using partitions of the unity.

In such a case, we obtain by restriction a connection $\bar{\Gamma}$ in
the fibre bundle $\pi_{X W_1} : W_1 \longrightarrow X$, which is a solution of
equation (\ref{connection}) when it is restricted to $W_1$
(in fact, we have a family of such solutions).
Assume that $\bar{\Gamma}$ is flat, and $\bar{\psi}$ is a horizontal
section of $\bar{\Gamma}$. First of all, notice that $\bar{\psi}$ takes values
in $W_1$ which implies that $\psi = \hbox{pr}_2 \circ \bar{\psi}$ is a jet prolongation.
Let us explain better this assertion.
If $\bar{\psi}(x^\mu) = (x^{\mu}, y^i (x), p(x), p^{\mu}_i(x), z^{i}_{\mu}(x))$
then we have
$$
z^{i}_{\mu}(x) = \frac{\partial y^i}{\partial x^\mu}.
$$
Since
$$
D^i_{\mu \nu} = \frac{\partial z^i_\nu}{\partial x^\mu}
$$
we deduce that along $\psi$ we have
\begin{eqnarray*}
\frac{\partial L}{\partial y^j}
-\frac{\partial^2 L}{\partial z^j_{\mu} \partial x^{\mu}}
-\frac{\partial y^i}{\partial x^\mu} \frac{\partial^2 L}{\partial z^j_{\mu} \partial y^{i}}
-\frac{\partial z^i_\nu}{\partial x^\mu}\frac{\partial^2 L}{\partial z^j_{\mu} \partial z^{i}_{\nu}}=0
\end{eqnarray*}
that is,
\[
\frac{\partial L}{\partial y^j}-\frac{d}{dx^{\mu}}\left(\frac{\partial L}{\partial z^j_{\mu}}\right)=0
\]
which are the Euler-Lagrange equations for $L$.

Up to now, we have no assigned any meaning to the coordinate $p$.
Consider the submanifold $\bar{W}_1$ of $W_1$
defined by the equation $H_0 = 0$.
In other words, $\bar{W}_1$ is locally characterized by the equation
$$
p = - (p^\mu_i z^i_\mu - L),
$$
which defines a local energy.

We can ask when a solution exists on $\bar{W}_1$.
Indeed, it is possible to construct a family of connections
in the fibre bundle $\pi_{X\bar{W}_1} : \bar{W}_1 \longrightarrow X$
which solve equation (\ref{connection}) as follows.

We have to choose coefficients $B_{\mu}$, $C^{\nu}_{\mu i}$, and
$D^{i}_{\mu \nu}$ verifying (\ref{dosmas}) and (\ref{matrices1}), and
in addition,
\begin{equation}\label{barra1}
\bar{h}(\frac{\partial}{\partial x^\mu})(H_0)=0.
\end{equation}
A direct computation shows that (\ref{barra1}) is equivalent to the
following local conditions
\begin{equation}\label{barra2}
B_\mu + C^{\nu}_{\mu i} z^i_\nu =
\frac{\partial L}{\partial x^\mu} + z^i_\mu \frac{\partial L}{\partial y^i}.
\end{equation}
Now, if we choose appropriate values for $C^\nu_{\mu i}$
satisfying (\ref{dosmas}) and (\ref{matrices1}), then we can take
the values for $B_\mu$ given by equation (\ref{barra2}).
A global solution is finally obtained using partitions of the unity.

\medskip

Denote by $\Omega_{\bar{W}_1}$ the restriction of $\Omega_{H_0}$ to $\bar{W}_1$.

\th{Proposition}{5.1.
If $n \geq 2$ and the Lagrangian $L$ is regular then
$\Omega_{\bar{W}_1}$ is a multisymplectic form.
}
\Proof
The result follows from a direct computation taking into account that
on $W_1$ we have
\[
p_i^{\mu} = \frac{\partial L}{\partial z^i_{\mu}}
\]
and that the hessian matrix
$$
\left(\frac{\partial^2L}{\partial z^i_\mu \partial z^j_\nu}\right)
$$
is regular.
\sq

\vskip4pt plus2pt

\medskip

Next, we shall relate the above construction with the precedent ones
on the lagrangian and the hamiltonian sides.

First of all, the following results are quite obvious:

\begin{itemize}

\item The submanifold $\bar{W}_1$ is diffeomorphic to $Z$.

\item If $n \geq2$ and $L$ is (hyper)regular, then the
multisymplectic manifolds  $(\bar{W}_1, \Omega_{\bar{W}_1})$, $(Z,
\Omega_L)$ and $(Z^*, \Omega_h)$ are (globally) locally
multisymplectomorphic. Indeed, the corresponding
multisymplectomorphisms are the following ones:
\begin{eqnarray*}
&& (\hbox{pr}_2)_{|_{\bar{W}_1}} : \bar{W}_1 \longrightarrow Z\\
&& Leg_L : Z \longrightarrow Z^*\\
&& Leg_L \circ (\hbox{pr}_2)_{|_{\bar{W}_1}} : \bar{W}_1 \longrightarrow Z^*
\end{eqnarray*}
(Note that $\lambda \circ (\hbox{pr}_1)_{|_{\bar{W}_1}}
= Leg_L \circ (\hbox{pr}_2)_{|_{\bar{W}_1}}$).
\item As a consequence, one can choose
connections ${\bf h}$, $\tilde{\bf h}$ and $\bar{\bf h}$ in
the fibrations $\pi_{XZ}: Z \longrightarrow X$,
$\pi_{X Z^*}: Z^* \longrightarrow X$,
and $\pi_{X \bar{W}_1}: \bar{W}_1 \longrightarrow X$, respectively, such that they are
solutions of equations
(\ref{connection1}), (\ref{connection2}) and (\ref{connection}), respectively, and,
in addition, they are related by the above multisymplectomorphisms.
\end{itemize}

The following diagram summarizes the above discussion:

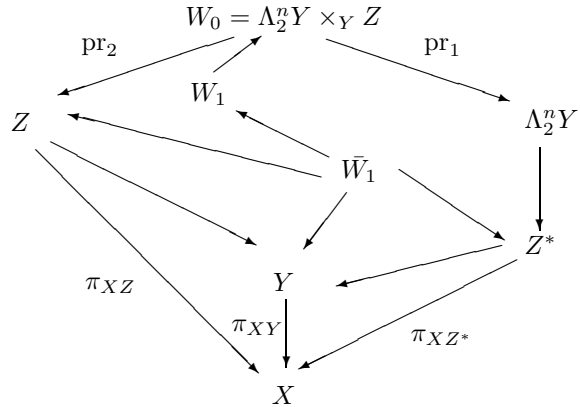
\begin{center}
\unitlength=1mm
\special{em:linewidth 0.4pt}
\linethickness{0.4pt}
\begin{picture}(100,80.00)(25,0)
\put(80.00,70.00){\makebox(0,0)[cc]{$W_0=\Lambda_2^n Y\times_Y Z$}}
\put(73.33,67.67){\vector(-3,-1){23.33}}
\put(85.67,67.33){\vector(3,-1){24.33}}
\put(46.33,56.33){\makebox(0,0)[rc]{$Z$}}
\put(112.00,56.33){\makebox(0,0)[lc]{$\Lambda_2^n Y$}}
\put(58.00,66.00){\makebox(0,0)[rb]{$\hbox{pr}_2$}}
\put(98.67,66.00){\makebox(0,0)[lb]{$\hbox{pr}_1$}}
\put(70.00,60.00){\makebox(0,0)[cc]{$W_1$}}
\put(90.00,50.00){\makebox(0,0)[cc]{$\bar{W_1}$}}
\put(80.00,35.00){\makebox(0,0)[cc]{$Y$}}
\put(80.00,20.00){\makebox(0,0)[cc]{$X$}}
\put(70.67,63.00){\vector(4,3){6.33}}
\put(86.00,51.67){\vector(-2,1){12.33}}
\put(85.00,49.00){\vector(-4,1){33.67}}
\put(88.33,47.00){\vector(-3,-4){5.67}}
\put(114,53){\vector(0,-1){11}}
\put(49.00,53.67){\vector(2,-1){27.33}}
\put(112.00,40.00){\makebox(0,0)[lc]{$Z^*$}}
\put(95.33,50){\vector(3,-2){14}}
\put(109.00,40){\vector(-4,-1){22.00}}
\put(47.00,52.67){\vector(1,-1){29.67}}
\put(111, 38){\vector(-2,-1){29.00}}
\put(60.33,35.67){\makebox(0,0)[rt]{$\pi_{XZ}$}}
\put(80.33,32.67){\vector(0,-1){9.00}}
\put(80,29.00){\makebox(0,0)[rc]{$\pi_{XY}$}}
\put(101.00,27.67){\makebox(0,0)[cc]{$\pi_{XZ^*}$}}
\end{picture}
\end{center}
\vspace{-2cm}

\section*{6. Singular lagrangians.}\label{singularsection}
For a singular lagrangian $L$, we usually have to go further in the constraint
algorithm. Therefore, we will consider a subset $\bar{W}_2$ defined in order to satisfy
the tangency condition:
\begin{eqnarray*}
\bar{W}_2 & = & \{u \in \bar{W}_1 \; /\; \exists  \bar{\hbox{\bf h}}_{u} : T_u W_0
\longrightarrow T_u \bar{W}_1 \quad \hbox{linear such that }\
\bar{\hbox{\bf h}}^2_u = \bar{\hbox{\bf h}}_u,\\
&&\ker \bar{\hbox{\bf h}}_u = (V\pi_{X W_0})_u, \ i_{\bar{\hbox{\bf h}}_u}
\Omega_{H_0}(u) = (n-1) \Omega_{H_0}(u)\}.
\end{eqnarray*}
Assume that $\bar{W}_2$ is a submanifold of $\bar{W}_1$.
If $\bar{\hbox{\bf h}}_u(T_uW_0)$ is not contained in $T_u\bar{W}_2$,
we go to the third step, and so on.

At the end, and if the system has solutions, we will find a final
constraint submanifold $\bar{W}_f$, fibered over $X$ (or over some
open subset of $X$) (see Appendix C) and a connection
$\bar{\Gamma}_f$ in this fibration such that $\bar{\Gamma}_f$ is a
solution of equation (\ref{connection}) restricted to $\bar{W}_f$.

Similar constraint algorithms can be developed using equations
(\ref{connection1}) and (\ref{connection2}). Our purpose in the following is
to relate these three algorithms.

\medskip

Indeed, we can consider the subset
\begin{eqnarray*}
Z_2 & = & \{z \in Z \; /\; \exists  \hbox{\bf h}_{z} : T_z Z
\longrightarrow T_zZ \quad \hbox{linear such that }\
\hbox{\bf h}^2_z = \hbox{\bf h}_z,\\
&&\ker \hbox{\bf h}_z = (V\pi_{X Z})_z,  i_{\hbox{\bf h}_z}
\Omega_{L}(z) = (n-1) \Omega_{L}(z)\}.
\end{eqnarray*}
If $Z_2$ is a submanifold, then there are solutions but we have to include the tangency
condition, and consider a new step:
\begin{eqnarray*}
Z_3 & = & \{z \in Z_2 \; /\; \exists  \hbox{\bf h}_{z} : T_zZ
\longrightarrow T_z Z_2 \quad \hbox{linear such that }\
\hbox{\bf h}^2_z = \hbox{\bf h}_z,\\
&&\ker \hbox{\bf h}_z = (V\pi_{X Z})_z, \, i_{\hbox{\bf h}_z}
\Omega_{L}(z) = (n-1) \Omega_{L}(z)\}.
\end{eqnarray*}
If $Z_3$ is a submanifold of $Z_2$, but $\hbox{\bf h}_z(T_zZ)$ is not contained in $T_zZ_3$,
we go to the third step, and so on.
Finally, we will obtain (in the favorable cases) a final
constraint submanifold $Z_f$ and a connection in the fibration
$\pi_{XZ}: Z \longrightarrow X$ along the
submanifold $Z_f$ (in fact, a family of connections)
with horizontal projector $\hbox{\bf h}$
which is a solution of equation (\ref{connection1}).

There is an additional problem, since our connection would
be a solution of the de Donder problem,
but not a solution of the Euler-Lagrange equations.
This problem is solved constructing
a submanifold of $Z_f$ where such a solution exists
(see \cite{LMMS,LMMS2} and below for more details).

\medskip

To develop a hamiltonian counterpart, we need some weak regularity of the lagrangian $L$.

\defin{Definition}
{6. 1. A lagrangian $L :  Z\longrightarrow \R$ is said to be almost regular if
$leg_L(Z)=\tilde{Z}$ is
a submanifold of $\Lambda^n_2Y$, and $leg_L : Z \longrightarrow \tilde{Z}$
is a submersion with connected fibers.
}

If $L$ is almost regular, one has:

\begin{itemize}

\item $\tilde{Z}_1 = Leg_L(Z)$ is a submanifold of $Z^*$, and in
addition, a fibration over $X$.

\item The restriction $\lambda_1 : \tilde{Z} \longrightarrow \tilde{Z}_1$ of $\lambda$
is a diffeomorphism.

\item The mapping $Leg_1 : Z \longrightarrow \tilde{Z}_1$ is a submersion
with connected fibers.
\end{itemize}

Define a mapping $h_1=(\lambda_1)^{-1} : \tilde{Z}_1 \longrightarrow \tilde{Z}$,
and a $(n+1)$-form $\tilde{\Omega}_1$ on $\tilde{Z}_1$ by
$\tilde{\Omega}_1 = h_1^* ((\Omega_2)_{_{|_{\tilde{Z}}}})$.
Obviously, we have $Leg_1^* \tilde{\Omega}_1 = \Omega_L$.

The hamiltonian description is now based in the equation
\begin{equation}\label{connection3}
i_{\tilde{\hbox{\bf h}}} \tilde{\Omega}_{1} = (n-1)
\tilde{\Omega}_{1}
\end{equation}
where $\tilde{\hbox{\bf h}}$ is a connection in the fibration
$\pi_{X \tilde{Z}_1} : \tilde{Z}_1 \longrightarrow X$.

Proceeding as above, we construct a constraint algorithm as follows.

First, we define
\begin{eqnarray*}
\tilde{Z}_2 & = & \{\tilde{z} \in \tilde{Z}_1 \; /\; \exists
\tilde{\hbox{\bf h}}_{\tilde{z}} : T_{\tilde{z}} \tilde{Z}_1
\longrightarrow T_{\tilde{z}}\tilde{Z}_1 \quad \hbox{linear such that }\
\tilde{\hbox{\bf h}}^2_{\tilde{z}} = \tilde{\hbox{\bf h}}_{\tilde{z}},\\
&&\ker \tilde{\hbox{\bf h}}_{\tilde{z}} = (V\pi_{X \tilde{Z}_1})_{\tilde{z}},
i_{\tilde{\hbox{\bf h}}_{\tilde{z}}}
\tilde{\Omega}_{1}(\tilde{z}) = (n-1) \tilde{\Omega}_{1}(\tilde{z})\}.
\end{eqnarray*}
If $\tilde{Z}_2$ is a submanifold, then there are solutions but
we have to include the tangency
condition, and consider a new step:
\begin{eqnarray*}
\tilde{Z}_3 & = & \{\tilde{z} \in \tilde{Z}_2 \; /\; \exists  \tilde{\hbox{\bf h}}_{\tilde{z}}
: T_{\tilde{z}}\tilde{Z}_1
\longrightarrow T_{\tilde{z}} \tilde{Z}_2 \quad \hbox{linear such that }\
\tilde{\hbox{\bf h}}^2_{\tilde{z}} = \tilde{\hbox{\bf h}}_{\tilde{z}},\\
&&\ker \tilde{\hbox{\bf h}}_{\tilde{z}} = (V\pi_{X \tilde{Z}_{1}})_{\tilde{z}},\,
i_{\tilde{\hbox{\bf h}}_{\tilde{z}}}
\tilde{\Omega}_{1}(\tilde{z}) = (n-1) \tilde{\Omega}_{1}(\tilde{z})\}.
\end{eqnarray*}
If $\tilde{Z}_3$ is a submanifold of $\tilde{Z}_2$, but
$\tilde{\hbox{\bf h}}_{\tilde{z}}(T_{\tilde{z}}\tilde{Z}_1)$ is not contained in
$T_{\tilde{z}}\tilde{Z}_3$, we go to the third step, and so on.
Finally, we will obtain (in the favorable cases) a final
constraint submanifold $\tilde{Z}_f$ and a connection in the fibration
$\pi_{X\tilde{Z}_1} : \tilde{Z}_{1} \longrightarrow
X$ along the submanifold $\tilde{Z}_f$
(in fact, a family of connections)
with horizontal projector $\tilde{\hbox{\bf h}}$
which is a solution of equation (\ref{connection2}).

\medskip

The important facts are the following:

\begin{itemize}

\item The mapping $Leg_1 : Z \longrightarrow \tilde{Z}_1$ preserves
the constraint algorithms, that is, we have
$Leg_1(Z_r)=\tilde{Z}_r$ for each integer $r \geq 2$.

\item In consequence, both algorithms have the same behavior; in particular,
if one of them stabilizes, the same happens with the other, and at
the same step, so we have $Leg_1(Z_f)=\tilde{Z}_f$.

\item In the latter case, the restriction $Leg_f : Z_f \longrightarrow \tilde{Z}_f$
is a surjective submersion (that is, a fibration) and
$Leg^{-1}_{f}(Leg_{f}(z)) = Leg_{1}^{-1}(Leg_{1}(z))$, for all $z \in
Z_{f}$.

\end{itemize}

Therefore, the lagrangian and hamiltonian sides can be compared
through the fibration $Leg_f :Z_f \longrightarrow \tilde{Z}_f$.
Indeed, if we have a connection in the fibration
$\pi_{XZ} : Z \longrightarrow X$ along the submanifold $Z_f$
with horizontal projector  $\hbox{\bf h}$ which is a solution
of equation (\ref{connection1}) (the de Donder equation)
and, in addition, the connection is projectable via $Leg_f$ to
a connection in the fibration $\pi_{X \tilde{Z}} : \tilde{Z} \longrightarrow X$
along the submanifold $\tilde{Z}_f$, then the
horizontal projector of the projected connection is a solution
of equation (\ref{connection2}) (the Hamilton equations).
Conversely, given a connection in the fibration
$\pi_{X \tilde{Z}} : \tilde{Z} \longrightarrow X$ along
the submanifold $\tilde{Z}_f$, with horizontal projector
$\tilde{\hbox{\bf h}}$ which is a solution of equation (\ref{connection2}),
then every connection in the fibration
$\pi_{XZ} : Z \longrightarrow X$ along the submanifold $Z_f$
that projects onto $\tilde{\hbox{\bf h}}$ is a solution of the de
Donder equation (\ref{connection1}).

\medskip

Assume that $L$ is almost regular and construct the above
algorithms. Take a $Leg_f$-projectable connection
$\Gamma$ in the fibration
$\pi_{XZ} : Z \longrightarrow X$ along the submanifold $Z_f$
with horizontal projector $\hbox{\bf h}$ which is a solution
of equation (\ref{connection1}), and denote by $\tilde{\Gamma}$ its projection.
As we have shown, the horizontal projector $\tilde{\hbox{\bf h}}$
is a solution of equation (\ref{connection2}).

In general, $\Gamma$ is not semi-holonomic, that is, $S_\eta(\hbox{\bf h},
\dots, \hbox{\bf h}) \not\equiv 0$ along $Z_f$.
However, we can define a section $\beta$ of the fibration
$Leg_L : Z_f \longrightarrow \tilde{Z}_f$ such that
$$
(S_\eta(\hbox{\bf h}, \dots, \hbox{\bf h}))_{|_{\beta(\tilde{Z}_{f})}} = 0.
$$
The construction of $\beta$
is based in the following interpretation of the elements of $Z$.

Take $z \in Z$, that is, $z$ is a 1-jet of a section
$\phi$ of the fibration $\pi_{XY} : Y \longrightarrow X$.
Since $\hbox{\bf H}_{\phi(x)}=T\phi(x)(T_xX)$ is a horizontal subspace of $T_{\phi(x)}Y$,
for every $x \in X$ (in fact, in the domain of $\phi$)
we can identify $z$ with this horizontal subspace, which in local
coordinates means that if $z = (x^\mu, y^i, z^i_\mu)$, then
$\hbox{\bf H}_{\phi(x)}$ is spanned by the tangent vectors
$\displaystyle{\frac{\partial}{\partial x^\mu} + z^i_\mu \frac{\partial}{\partial y^i}}$.

With the above notations and the obvious identifications, we define
\begin{equation}\label{seccion}
\beta(\tilde{z}) = T\pi_{YZ}({\bf h}(T_{z_0}Z)),
\end{equation}
where $z_0 \in Z_f$ is an arbitrary point projecting onto $\tilde{z}$ through
the projection $Leg_f : Z_f \longrightarrow \tilde{Z}_f$.

We have:

\begin{itemize}

\item $\beta(\tilde{z})$ is independent of the choice of $z_0$. This is a consequence
of the following two facts: $\hbox{\bf h}$ projects onto $\tilde{\hbox{\bf h}}$, and
the relation $\pi_{XZ^*} \circ Leg_f = \pi_{XZ}$.

\item The point $\beta(\tilde{z})$ belongs to $Z_f$. Indeed,
consider the following local vector field
$$
U = (\Gamma^i_\mu - z^i_\mu) \frac{\partial}{\partial z^i_\mu},
$$
where $\Gamma^i_\mu$ are the Christoffel components of $\Gamma$, that is
$$
\hbox{\bf h}(\frac{\partial}{\partial x^\mu})=
\frac{\partial}{\partial x^\mu} + \Gamma^i_\mu \frac{\partial}{\partial y^i}
+ \Gamma^i_{\mu \nu} \frac{\partial}{\partial z^i_\nu}.
$$
Since $\Gamma$ is $Leg_f$-projectable, then $\Gamma^i_\mu$ is constant along the
fibre over $\tilde{z}$.

>From (\ref{uno}) and (\ref{legendre}), we deduce that $U$ is a
vertical vector field with respect to the fibration $Leg_f : Z_f
\longrightarrow \tilde{Z}_f$, and in consequence it is tangent to
the fibre over $\tilde{z}$. Consider the curve $$ \alpha(t) =
((x^\mu)_0, (y^i)_0, (\Gamma^i_\mu)_0 - \exp (-t)
((\Gamma^i_\mu)_0 - (z^i_\mu)_{0})), $$ where $((x^\mu)_0,
(y^i)_0, (z^i_\mu)_0)$ are the coordinates of $z_0$, and
$(\Gamma^i_\mu)_{0}$ are the values of $\Gamma^i_\mu$ at the point
$z_0$ (in fact, along all the fibre). $\alpha (t)$ is an integral
curve of $U$ passing through $z_0$ and totally contained in the
fibre over $\tilde{z}$. Thus, the limit point $\lim_{t \to
+\infty} \, \alpha(t)$ is in this fibre, and a direct computation
shows that $\lim_{t \to +\infty} \, \alpha(t) = \beta(\tilde{z})$.

\item Now, it is obvious that $\Gamma$ is semiholonomic at the point $\beta(\tilde{z})$.

\end{itemize}

Since $\beta$ is a section, we deduce that $\beta(\tilde{Z}_f)$ is a submanifold of
$Z_f$ and hence of $Z$. In addition,
$(Leg_f)_{|_{\beta(\tilde{Z}_f)}} : \beta(\tilde{Z}_f)
\longrightarrow \tilde{Z}_f$ is a diffeomorphism.

Next, we define a connection $\Gamma_s$ in the fibration
$\pi_{XZ} : Z \longrightarrow X$ along $\beta(\tilde{Z}_f)$ as follows.

Its horizontal projector is given by
$$
(\hbox{\bf h}_s)_z : T_zZ \longrightarrow T_z \beta(\tilde{Z}_f), \quad
(\hbox{\bf h}_s)_z =
(T(Leg_f)_{|_{\beta(\tilde{Z}_f)}}(z))^{-1} \circ
\tilde{\hbox{\bf h}}_{\tilde{z}} \circ
TLeg_f(z),
$$
for all $z \in \beta(\tilde{Z}_f)$, where $z = \beta(\tilde{z})$.
A straightforward computation shows that $\Gamma_s$ is a solution
of (\ref{connection1}) and, in addition, is transported onto $\tilde{\Gamma}$
via the diffeomorphism $(Leg_L)_{|_{\beta(\tilde{Z}_f)}}
: \beta(\tilde{Z}_f) \longrightarrow \tilde{Z}_f$. Thus, since
$\Gamma$ is semiholonomic along $\beta(\tilde{Z}_f)$, we deduce that
$\Gamma_{s}$ is also semiholonomic along $\beta(\tilde{Z}_f)$.

\medskip

Next, we will relate the above constructions with the algorithm
developed from equation (\ref{connection}).

To do that, we first develop an alternative constraint algorithm based
in the following equation

\begin{equation}\label{alternative}
i_{\hat{\hbox{\bf h}}} \, \Omega_{\bar{W}_1} = (n-1) \Omega_{\bar{W}_1},
\end{equation}
where $\Omega_{\bar{W}_1}$ is the restriction of $\Omega_{H_0}$ to
$\bar{W}_1$, and $\hat{\hbox{\bf h}}$ is the horizontal projector of
a connection $\hat{\Gamma}$ in the fibration $\pi_{X
\bar{W}_1}=(\pi_{XW_0})_{|_{\bar{W}_1}} : \bar{W}_1 \longrightarrow
X$.

The algorithm proceed now as in the above cases, and it
produces a chain of submanifolds (in the favorable cases).
Indeed, we define

\begin{eqnarray*}
\hat{W}_2 & = & \{u \in \bar{W}_1 \; /\; \exists  \hat{\hbox{\bf h}}_{u} : T_u \bar{W}_1
\longrightarrow T_u \bar{W}_1 \quad \hbox{linear such that }\
\hat{\hbox{\bf h}}^2_u = \hat{\hbox{\bf h}}_u,\\
&&\ker \hat{\hbox{\bf h}}_u = (V\pi_{X \bar{W}_1})_u, \
i_{\hat{\hbox{\bf h}}_u} \Omega_{\bar{W}_1}(u) = (n-1) \Omega_{\bar{W}_1}(u)\}.
\end{eqnarray*}
If we assume that $\hat{W}_2$ is a submanifold of $\bar{W}_1$, since
in general $\hat{\hbox{\bf h}}_u(T_u\bar{W}_1)$ is not contained in $T_u\hat{W}_2$,
we go to the third step, and so on.

At the end, and if the system has solutions, we will find a final
constraint submanifold $\hat{W}_f$, fibered over $X$ (or over some
open subset of $X$) (see Appendix C) and a connection
$\hat{\Gamma}_f$ in this fibration such that $\hat{\Gamma}_f$ is a
solution of equation (\ref{alternative}) restricted to
$\hat{W}_f$.

It should be noticed that $\bar{W}_r \subset \hat{W}_r$,
for all integer $r \geq 2$. Indeed, any pointwise solution of
equation (\ref{connection}) is a solution of equation (\ref{alternative}).
As a consequence, both algorithms have the same behavior.

\medskip

This last algorithm can be compared with the lagrangian
and hamiltonian ones. In fact, since
$$
\tilde{\hbox{pr}}_2^* \Omega_L = \Omega_{\bar{W}_1} , \quad
(\tilde{\hbox{pr}}_1)^* \tilde{\Omega}_{1} = \Omega_{\bar{W}_1},
$$
where
$\tilde{\hbox{pr}}_1 = \lambda_1 \circ (\hbox{pr}_1)_{|_{\bar{W}_1}}$ and
$\tilde{\hbox{pr}}_2 = (\hbox{pr}_2)_{|_{\bar{W}_1}}$, we have
$$
\tilde{\hbox{pr}}_1(\hat{W}_r) = \tilde{Z}_r, \quad
\tilde{\hbox{pr}}_2 (\hat{W}_r) = Z_r,
$$
for all $r \geq 2$, and {\it a fortiori} we deduce that
all the algorithms have the same behavior and
$$
\tilde{\hbox{pr}}_1(\hat{W}_f) = \tilde{Z}_f, \quad
\tilde{\hbox{pr}}_2 (\hat{W}_f) = Z_f.
$$
Thus, the corresponding solutions can be related via
the convenient projections. More precisely,
we can construct a connection $\Gamma$ (resp. $\tilde{\Gamma}$,
$\hat{\Gamma}$) in the fibration
$\pi_{XZ} : Z \longrightarrow X$
(resp. $\pi_{X\tilde{Z}_1} : \tilde{Z}_1 \longrightarrow X$,
$\pi_{X\bar{W}_1} : \bar{W}_1 \longrightarrow X$)
along the submanifold $Z_f$ (resp. $\tilde{Z}_f$, $\hat{W}_f$)
such that they are related by the projections
$Leg_f$, $\tilde{\hbox{pr}}_1$ and $\tilde{\hbox{pr}}_2$.

In addition, the connection $\Gamma$ can be chosen such that its
restriction to $\bar{W}_f$ is a solution of equation (\ref{connection}).
Making all these selections, and performing the construction of
the section $\beta$ we conclude that
$\beta(\tilde{Z}_f) \subset \bar{W}_f$.

The following diagram summarizes the above discussion:

\bigskip

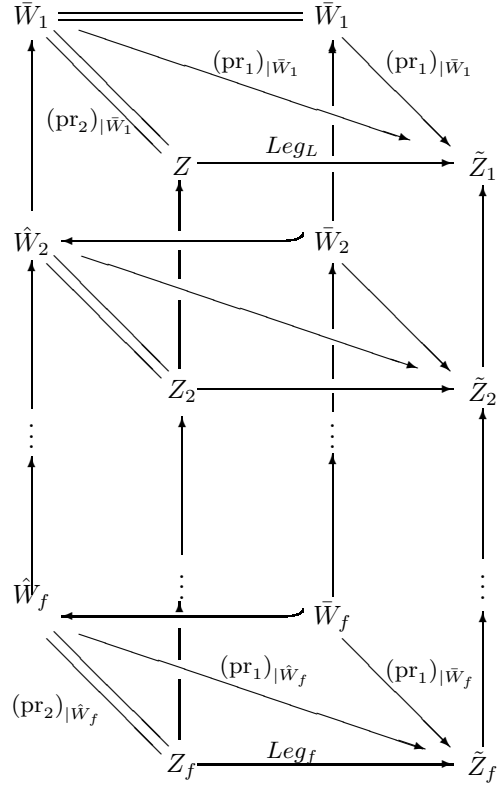
\unitlength=1.00mm
\special{em:linewidth 0.4pt}
\linethickness{0.4pt}
\begin{picture}(100.00,100.33)
\put(40.00,100.00){\makebox(0,0)[cc]{$\bar{W}_1$}}
\put(80.00,100.00){\makebox(0,0)[cc]{$\bar{W}_1$}}
\put(60.00,80.00){\makebox(0,0)[cc]{$Z$}}
\put(100.00,80.00){\makebox(0,0)[cc]{$\tilde{Z}_1$}}
\put(42.00,97.00){\line(1,-1){15.25}}
\put(43.00,98.00){\line(1,-1){15.25}}
\put(43.63,100.33){\line(1,0){32.33}}
\put(43.63,99.33){\line(1,0){32.33}}
\put(81.33,97.00){\vector(1,-1){14.33}}
\put(62.00,80.33){\vector(1,0){34.00}}
\put(74.67,81.67){\makebox(0,0)[cb]{\small $Leg_L$}}
\put(53.73,87.67){\makebox(0,0)[rt]{\small $(\hbox{pr}_2)_{|\bar{W}_1}$}}
\put(87,93.00){\makebox(0,0)[lb]{\small $(\hbox{pr}_1)_{|\bar{W}_1}$}}
\put(64.00,93.00){\makebox(0,0)[lb]{\small $(\hbox{pr}_1)_{|\bar{W}_1}$}}
\put(40.00,70.00){\makebox(0,0)[cc]{$\hat{W}_2$}}
\put(80.00,70.00){\makebox(0,0)[cc]{$\bar{W}_2$}}
\put(60.00,50.00){\makebox(0,0)[cc]{$Z_2$}}
\put(100.00,50.00){\makebox(0,0)[cc]{$\tilde{Z}_2$}}
\put(42.00,67.00){\line(1,-1){15.25}}
\put(43.00,68.00){\line(1,-1){15.25}}
\put(81.33,67.00){\vector(1,-1){14.33}}
\put(62.00,50.33){\vector(1,0){34.00}}
\put(40.00,45.00){\makebox(0,0)[cc]{$\vdots$}}
\put(80.00,45.00){\makebox(0,0)[cc]{$\vdots$}}
\put(60.00,25.00){\makebox(0,0)[cc]{$\vdots$}}
\put(100.00,25.00){\makebox(0,0)[cc]{$\vdots$}}
\put(40.00,23.00){\makebox(0,0)[cc]{$\hat{W}_f$}}
\put(80.00,20.00){\makebox(0,0)[cc]{$\bar{W}_f$}}
\put(60.00,0.00){\makebox(0,0)[cc]{$Z_f$}}
\put(100.00,0.00){\makebox(0,0)[cc]{$\tilde{Z}_f$}}
\put(42.00,17.00){\line(1,-1){15.25}}
\put(43.00,18.00){\line(1,-1){15.25}}
\put(73.00,20.00){\vector(-1,0){29.00}}
\bezier{40}(73.00,20.00)(76,20.00)(76,21.00)
\put(81.33,17.00){\vector(1,-1){14.33}}
\put(62.00,0.33){\vector(1,0){34.00}}
\put(74.67,1.2){\makebox(0,0)[cb]{\small $Leg_f$}}
\put(49.33,9.67){\makebox(0,0)[rt]{\small $(\hbox{pr}_2)_{|\hat{W}_f}$}}
\put(87.00,12.00){\makebox(0,0)[lb]{\small $(\hbox{pr}_1)_{|\bar{W}_f}$}}
\put(65,12.5){\makebox(0,0)[lb]{\small $(\hbox{pr}_1)_{|\hat{W}_f}$}}
\put(73.00,70.00){\vector(-1,0){29.00}}
\bezier{40}(73.00,70.00)(76,70.00)(76,71.00)
\put(46.33,98.00){\vector(3,-1){44.00}}
\put(46.67,68.00){\vector(3,-1){44.67}}
\put(40.00,74.00){\vector(0,1){22.67}}
\put(59.67,53.00){\line(0,1){9.00}}
\put(59.67,65.00){\line(0,1){3.67}}
\put(59.67,71.33){\vector(0,1){6.67}}
\put(80.00,72.67){\line(0,1){6.33}}
\put(80.00,81.33){\line(0,1){4.00}}
\put(80.00,96.33){\vector(0,1){0.33}}
\put(80.00,88.00){\vector(0,1){8.33}}
\put(100.00,53.33){\vector(0,1){24.00}}
\put(40.00,47.67){\vector(0,1){19.67}}
\put(60.00,28.33){\vector(0,1){18.33}}
\put(80.00,47.33){\line(0,1){2.33}}
\put(80.00,51.67){\line(0,1){4.33}}
\put(80.00,58.33){\vector(0,1){8.67}}
\put(100.00,27.67){\vector(0,1){20.00}}
\put(40.00,23.00){\vector(0,1){18.33}}
\put(46.33,17.67){\vector(3,-1){46.00}}
\put(59.67,3.67){\line(0,1){8.00}}
\put(59.67,14.33){\line(0,1){4.00}}
\put(59.67,21.00){\vector(0,1){1.00}}
\put(80.00,22.67){\vector(0,1){19.00}}
\put(100.00,2.67){\vector(0,1){18.00}}
\end{picture}

\remar{Remark}{6.2.
According to Appendix C, one has that all the connections
considered in this section define {\it bona fide} connections
in the corresponding restricted fibrations
\begin{eqnarray*}
\pi_{X_0 Z_f} & : & Z_f \longrightarrow X_0, \\
\pi_{X_0 \tilde{Z}_f} & : & \tilde{Z}_f \longrightarrow X_0, \\
\pi_{X_0 \bar{W}_f} & : & \bar{W}_f \longrightarrow X_0, \\
\pi_{X_0 \hat{W}_f} & : & \hat{W}_f \longrightarrow X_0,
\end{eqnarray*}
where $X_0$ is an open submanifold of $X$.
}

\section*{7. Example: The bosonic string}{\rm (See \cite{BGP,gimmsy1})
Let $X$ be a 2-dimensional manifold, and $(B, g)$ a
$d+1$-dimensional spacetime
manifold endowed with a Lorentz metric $g$ of signature $(-,+,
\dots ,+)$.
A {\sl bosonic string} is a map $\phi : X
\longrightarrow B$.

 In the following, we will follow  the Polyakov approach to classical bosonic string theory. Let $S^{1,1}_2(X)$ be the bundle over $X$
of symmetric  2-covariant tensors with signature $(-,+)$ or $(1,1)$.
We take the vector bundle $\pi:Y=X\times B\times S^{1,1}_2(X)
\longrightarrow X$. Therefore,  in this formulation, a
 field $\psi$ is a section $(\phi, h)$ of the vector bundle $Y=X\times B\times S^{1,1}_2(X)\longrightarrow X$, where
$\phi : X\longrightarrow Y$ is the bosonic string
and $h$ is a Lorentz metric on $X$.

\medskip

{\bf Lagrangian description}

\medskip

We have that $Z= J^1(X\times B)\times_{X} J^1(S^{1,1}_2(X))$.
Taking coordinates $(x^{\mu})$, $(y^i)$ and $(x^{\mu},
h_{\mu\eta})$ on $X$, $B$ and $S^{1,1}_2(X)$ then the canonical
local coordinates on $Z$ are $(x^{\mu}, y^i
,h_{\eta\xi},$\linebreak  $y^i_{\mu},  h_{\eta\xi\mu})$. In this
system of local coordinates, the Lagrangian density  is given by
\[
\Lambda = -\frac{1}{2}\sqrt{-\det (h)}h^{\eta\xi}g_{ij}y^i_{\eta} y_{\xi}^jd^2 x\; .
\]
The Cartan 3-form is
\begin{eqnarray*}
\Omega_L&=& d y^i\wedge d\left( -\sqrt{-\det
(h)}h^{\eta\xi}g_{ij}y^j_{\xi} \right)\wedge d^1 x^{\eta}
         \\&&- d\left( \frac{1}{2}\sqrt{-\det
(h)}h^{\eta\xi}g_{ij}y^i_{\eta} y_{\xi}^j
\right)\wedge d^2x\\
&=& -\frac{1}{2}\left(\frac{\partial \sqrt{-\det (h)}}{\partial
h_{\rho\sigma}}h^{\eta\xi}g_{ij}y^i_{\eta}y^j_{\xi}-\sqrt{-\det
(h)}h^{\eta \rho}h^{\xi \sigma}g_{ij}y^{i}_{\eta}y^j_{\xi}\right)
dh_{\rho\sigma}\wedge d^2x\\
&&-\frac{1}{2}\sqrt{-\det (h)} h^{\eta\xi} \frac{\partial g_{ij}}{\partial y^k}
y^i_{\eta}y^j_{\xi}\,dy^k\wedge d^2x-\sqrt{-\det (h)} h^{\eta\xi}
g_{ij}y^i_{\eta}\, dy^j_{\xi}\wedge d^2x\\
&&+\left(  \frac{\partial \sqrt{-\det (h)}}{\partial
h_{\rho\sigma}}h^{\eta\xi} g_{ij}y^j_{\xi}-\sqrt{-\det (h)}h^{\eta
\rho}h^{\xi \sigma}g_{ij}y^j_{\xi} \right)d h_{\rho\sigma}\wedge
dy^i\wedge d^1 x^{\eta}\\
&&+\sqrt{-\det (h)} h^{\eta\xi} \frac{\partial g_{ij}}{\partial y^k}
y^j_{\xi}\, dy^k\wedge dy^i\wedge d^1 x^{\eta}\\&&+\sqrt{-\det (h)}
h^{\eta\xi} g_{ij}\, d y^j_{\xi}\wedge dy^i\wedge d^1x^{\eta}.
\end{eqnarray*}

If we solve the equation $i_{\bf h}\Omega_L=\Omega_L$, where
\[
{\bf h} = dx^{\mu}\otimes\left(\frac{\partial}{\partial x^\mu} +
{\Gamma}^i_{\mu}\frac{\partial}{\partial y^i} +
{\gamma}_{\eta\xi\mu}\frac{\partial}{\partial h_{\eta\xi}} +
{\Gamma}^i_{\eta\mu}\frac{\partial}{\partial y^i_{\eta}}
+{\gamma}_{\eta\xi\rho\mu}\frac{\partial}{\partial h_{\eta\xi\rho}}\right) \;,
\]
we obtain that:
\begin{eqnarray*}
\Gamma^i_{\mu}&=& y^i_{\mu}\\
0&=&\frac{1}{2}\sqrt{-\det (h)} h^{\eta\xi}\frac{\partial g_{ij}}{\partial y^k} y^i_{\eta}y^j_{\xi}
-\sqrt{-\det (h)} h^{\eta\xi}\frac{\partial g_{kj}}{\partial y^i} y^i_{\eta}y^j_{\xi}
-\sqrt{-\det (h)} h^{\eta\xi} g_{kj} \Gamma^j_{\xi\eta}\\&&-
\left(  \frac{\partial \sqrt{-\det (h)}}{\partial h_{\rho\sigma}}h^{\eta\xi} g_{kj}y^j_{\xi}-\sqrt{-\det (h)}h^{\eta \rho}h^{\xi \sigma}g_{kj}y^j_{\xi}
\right)\gamma_{\rho\sigma\eta}\;,
\end{eqnarray*}
and the constraints are given by the equations
\[
\frac{\partial }{\partial h_{\rho\theta}}\left(\sqrt{-\det (h)} h^{\eta\xi}\right)g_{ij} y^i_{\eta}y^j_{\xi}=0 \; .
\]
The previous equation corresponds to the three following constraints
\begin{eqnarray*}
\left[ h^{\eta 0}h^{\xi 0} (h_{01}^2-h_{00}h_{11}) + \frac{1}{2}
h^{\eta\xi}h_{11}\right] g_{ij} y^i_{\eta} y^j_{\xi} & = & 0 \\
\left[ h^{\eta 1}h^{\xi 1} (h_{01}^2-h_{00}h_{11}) + \frac{1}{2}
h^{\eta\xi}h_{00}\right] g_{ij} y^i_{\eta} y^j_{\xi} & = & 0 \\
\left[ h^{\eta 0}h^{\xi 1} (h_{01}^2-h_{00}h_{11}) -
h^{\eta\xi}h_{01}\right] g_{ij} y^i_{\eta} y^j_{\xi} & = & 0 \\
\end{eqnarray*}
which determine $Z_2$.

\medskip

\vspace{20pt}

{\bf Hamiltonian description}

\medskip

The Legendre transformation is given by
\[
Leg_L(x^{\mu}, y^i ,h_{\eta\xi},
y^i_{\mu}, h_{\eta\xi\mu})=(x^{\mu}, y^i ,h_{\eta\xi}, -\sqrt{-\det (h)}\, h^{\mu\eta}g_{ij}y^j_{\eta},
0)
\]
Therefore, the Lagrangian ${L}$ is almost-regular and, moreover,
$\tilde{Z}_1=\hbox{Im } Leg_L\cong \tilde{Z}=leg_L(Z)\cong
J^1(X\times B)\times_{X} S^{1,1}_2(X)$.
Take now  coordinates $(x^{\mu}, y^i ,h_{\eta\xi}, p_i^{\mu})$ on $\tilde{Z}_1$ and  consider the mapping $h_1: \tilde{Z}_1\rightarrow \tilde{Z}$ given by
\[
h_1 (x^{\mu}, y^i ,h_{\eta\xi}, p_i^{\mu})=(x^{\mu}, y^i ,h_{\eta\xi}, p=\frac{1}{2\sqrt{-\det(h)}}h_{\eta\xi}g^{ij}p^i_{\eta}p^j_{\xi}, p_i^{\mu})
\]

Then, we have
\[
\tilde{\Omega}_1=-d\left(\frac{1}{2\sqrt{-\det(h)}}h_{\eta\xi}g^{ij}p_i^{\eta}p_j^{\xi}\right)\wedge
d^2x+dy^i\wedge dp^{\mu}_i\wedge d^1x^{\mu}
\]
and the Hamilton equations are given by $i_{\tilde{\bf
h}}\tilde{\Omega}_1=\tilde{\Omega}_1$
\[
\tilde{\bf h}=
dx^{\mu}\otimes\left(\frac{\partial}{\partial x^\mu} +
\tilde{\Gamma}^i_{\mu}\frac{\partial}{\partial y^i} +
\tilde{\gamma}_{\eta\xi\mu}\frac{\partial}{\partial h_{\eta\xi}} +
\tilde{\Gamma}^{\eta}_{i\mu}\frac{\partial}{\partial p_i^{\eta}} \right)
\]
Solving the above equation, we obtain
\begin{eqnarray*}
\tilde{\Gamma}^i_{\mu}&=&-\frac{1}{\sqrt{-\det (h)}}h_{\eta\mu}g^{ij}p^{\eta}_j\\
\tilde{\Gamma}^{\mu}_{i\mu}&=&\frac{1}{2\sqrt{-\det (h)}}h_{\eta\xi}\frac{\partial g^{ij}}{\partial y^k} p^i_{\eta} p^j_{\xi} \;,
\end{eqnarray*}
and the secondary constraints
\[
\frac{g^{ij}}{\sqrt{-\det (h)}}\left(\frac{1}{2\det (h)}\frac{\partial \det (h)}{\partial h_{\rho\sigma}}h_{\eta\xi}p^{\eta}_ip^{\xi}_j-p^{\rho}_ip^{\sigma}_j\right)=0
\]
determining $\tilde{Z}_2$.

\medskip

{\bf The new geometrical setting}

\medskip

We have that $W_0=\Lambda^2_2 Y\times_{Y} Z$ with fibered coordinates
$$
(x^{\mu}, y^i ,h_{\eta\xi},
p, p^{\mu}_i, q^{\eta\xi\mu}, y^i_{\mu}, h_{\eta\xi\mu}).
$$
Therefore,
\begin{eqnarray*}
H_0&=&p+p^{\mu}_iy^i_{\mu}+q^{\eta\xi\mu}h_{\eta\xi\mu}+\frac{1}{2}\sqrt{-\det
(h)}h^{\eta\xi}g_{ij}y^i_{\eta} y_{\xi}^j, \\
\Omega_{H_0}&=&-dp\wedge d^2x-dp^{\mu}_i\wedge dy^i\wedge d^1 x^{\mu}-
dq^{\eta\xi\mu}\wedge d h_{\eta\xi}\wedge d^1 x^{\mu}+dH_0\wedge d^2x.\\
\end{eqnarray*}
Consider now an Ehresmann connection in the fibered manifold $\pi_{X W_0}: W_0
\longrightarrow X$ with horizontal projector:
\begin{eqnarray*}
\bar{\bf h}&=&
dx^{\mu}\otimes\left(\frac{\partial}{\partial x^\mu} +
A^i_{\mu}\frac{\partial}{\partial y^i} +
A_{\eta\xi\mu}\frac{\partial}{\partial h_{\eta\xi}} +
B_{\mu}\frac{\partial}{\partial p}
+C^{\eta}_{\mu i} \frac{\partial}{\partial p_i^{\eta}}
+C_{\eta\xi\sigma\mu}\frac{\partial}{\partial q^{\eta\xi\sigma}} \right.\\
&&\left.+D^{i}_{\eta\mu}\frac{\partial}{\partial y^i_{\eta}}
+D^{\eta\xi\sigma\mu}\frac{\partial}{\partial h_{\eta\xi\sigma}}
\right)
\end{eqnarray*}
Solving $i_{\bar{\bf h}}\Omega_{H_0}=\Omega_{H_0}$ we obtain that the
submanifold $W_1$ is determined by the constraints:
\begin{eqnarray*}
p^{\mu}_i&=&-\sqrt{-\det (h)} h^{\mu\eta}g_{ij}y^{j}_{\eta}\\
q^{\eta\xi\mu}&=&0
\end{eqnarray*}
Let $\bar{W}_1$ be the submanifold of $W_1$ defined by the equation $H_0=0$, that is
\[
p=\frac{1}{2}\sqrt{-\det (h)}h^{\eta\xi}g_{ij}y^i_{\eta} y_{\xi}^j.
\]
$\bar{W_1}$ is locally defined by coordinates $(x^{\mu}, y^i, h_{\eta\xi}, y^i_{\mu}, h_{\eta\xi\mu})$.

In this coordinates, the solutions of equation (\ref{alternative}) are exactly
the same than the ones obtained in the lagrangian setting, and $\hat{W}_2$,
as a submanifold of $W_0$, is determined by the vanishing of the constraints
functions
\begin{eqnarray*}
p^{\mu}_i+\sqrt{-\det (h)} h^{\mu\eta}g_{ij}y^{j}_{\eta}&=&0\\
q^{\eta\xi\mu}&=&0\\
p-\frac{1}{2}\sqrt{-\det (h)}h^{\eta\xi}g_{ij}y^i_{\eta} y_{\xi}^j&=&0\\
\frac{\partial \sqrt{-\det (h)}}{\partial h_{\rho\sigma}}h^{\eta\xi}g_{ij}y^i_{\eta}y^j_{\xi}-\sqrt{-\det (h)}h^{\eta \rho}h^{\xi \sigma}g_{ij}y^{i}_{\eta}y^j_{\xi}&=&0
\end{eqnarray*}
It is easy to show that $\bar{W}_2=\hat{W}_2$ and the solutions of equation
(\ref{connection}) are the solutions of equation (\ref{alternative}) which,
in addition, are semi-holonomic.
}

\section*{8. Time-dependent mechanics.}
The jet bundle description of time-dependent mechanical systems takes
$X = \R$ and $\eta = dt$, where $t$ is the usual coordinate on $\R$
(see, for instance, \cite{LMM}).

If $L: Z \longrightarrow \R$ is a lagrangian function, $\Omega_{L}$
is the Poincar\'e-Cartan $2$-form on $Z$ and $\eta_{Z}$ is the $1$-form
on $Z$ defined by $\eta_{Z}= (\pi_{{\R}Z})^{*}(\eta)$, then the de
Donder equation (\ref{connection1}) can be written as
\begin{equation}\label{time}
i_{\xi_{Z}}\Omega_{L} = 0, \makebox[.4cm]{} i_{\xi_{Z}}\eta_{Z} = 1,
\end{equation}
where $\xi_{Z}$ is a vector field on $Z$. The integral curves of
${\xi_{Z}}$ are the solutions of the de Donder problem.

The lagrangian function $L$ is regular if and only if the pair
$(\Omega_{L}, \eta_{Z})$ is a {\sl cosymplectic structure} on $Z$.
We recall that a cosymplectic structure on a manifold $M$ of odd
dimension $2n+1$ is a pair which consists of a closed $2$-form
$\Omega$ and a closed $1$-form $\eta$ such that $\eta \wedge
\Omega^{n}$ is a volume form.

If $L$ is regular then there exists a unique vector field $\xi_{Z}$
which satisfies (\ref{time}). In fact, $\xi_{Z}$ is the {\sl Reeb vector
field} of the cosymplectic structure $(\Omega_{L}, \eta_{Z})$ and it
is a second order differential equation, that is, $S_{dt} \xi_{Z} =
0$. The trajectories of $\xi_{Z}$ are the solutions of the
Euler-Lagrange equations.

On the other hand, in this case, $\Lambda^{1}_{2}Y$ is the cotangent
bundle $T^{*}Y$ of the manifold $Y$ and $\Omega_{0}$ is the canonical
symplectic structure of $T^{*}Y$. Moreover, if $h: Z^{*}
\longrightarrow \Lambda^{1}_{2}Y = T^{*}Y$ is a hamiltonian and
$\eta_{Z^{*}} = (\pi_{\R Z^{*}})^{*}(dt)$, then: i) the pair
$(\Omega_{h}, \eta_{Z^{*}})$ is a cosymplectic structure on $Z^{*}$ and
ii) the solutions of the Hamilton equations are just the integral
curves of the Reeb vector field $\xi_{h}$ of the cosymplectic
structure $(\Omega_{h}, \eta_{Z^{*}})$.

It should be noticed that if the lagrangian $L$ is regular and
$\eta_{\bar{W}_{1}} = \linebreak (\pi_{\R \bar{W}_{1}})^{*}(dt)$, we have that
the pair $(\Omega_{\bar{W}_{1}}, \eta_{\bar{W}_{1}})$ is again a
cosymplectic structure on $\bar{W}_{1}$ and there exists a unique
solution of equation (\ref{connection}) restricted to $\bar{W}_{1}$,
namely, the Reeb vector field of the cosymplectic structure
$(\Omega_{\bar{W}_{1}}, \eta_{\bar{W}_{1}})$. Furthermore, if $L$ is
(regular) hyper-regular then the maps $(\hbox{pr}_{2})_{|\bar{W}_{1}}:
\bar{W}_{1} \longrightarrow Z$, $Leg_{L}: Z \longrightarrow Z^{*}$
and $Leg_{L} \circ (\hbox{pr}_{2})_{|\bar{W}_{1}}: \bar{W}_{1}
\longrightarrow Z^{*}$ are (local) cosymplectomorphisms between the
cosymplectic manifolds $(\bar{W}_{1}, \Omega_{\bar{W}_{1}},
\eta_{\bar{W}_{1}})$, $(Z, \Omega_{L}, \eta_{Z})$ and $(Z^{*},
\Omega_{h}, \eta_{Z^{*}})$, where $h = leg_{L} \circ (Leg_{L})^{-1}$.
Thus, the Reeb vector fields $\xi_{\bar{W}_{1}}$, $\xi_{Z}$ and
$\xi_{Z^{*}}$ are related by the above cosymplectomorphisms.

When the lagrangian $L$ is singular, we can develop the two
algorithms using equations (\ref{connection}) and
(\ref{alternative}) and we obtain the corresponding constraint
submanifolds
\[\begin{array}{lclll}
\bar{W}_{i}&=&\{ u \in \bar{W}_{i-1} &/& \exists \xi \in T_{u}\bar{W}_{i-1},
\;i_{\xi}\Omega_{H_{0}}(u) = 0,\; \eta_{\bar{W}_{1}}(\xi) = 1 \},\\
\hat{W}_{i} &=& \{u \in \hat{W}_{i-1} &/& \exists \xi \in T_{u}\hat{W}_{i-1},
\;i_{\xi}\Omega_{\bar{W}_{1}}(u) = 0, \;\eta_{\bar{W}_{1}}(\xi) = 1 \},
\end{array}
\]
for all $i \geq 2$, with $\bar{W}_{1} = \hat{W}_{1}$ (see Section 6).

If $L$ is almost regular, then we have that
\[\begin{array}{lclll}
\bar{W}_{i}\subset \hat{W}_{i},\\
\tilde{\hbox{pr}}_1(\hat{W}_i) = \tilde{Z}_i& =& \{ \tilde{z} \in
\tilde{Z}_{i-1}& \kern-10pt/& \kern-10pt\exists\tilde{\xi} \in T_{\tilde{z}}\tilde{Z}_{i-1},\;
i_{\tilde{\xi}}\tilde{\Omega}_{1}(\tilde{z}) = 0,\;
\eta_{Z^{*}}(\tilde{z})(\tilde{\xi}) = 1 \},\\
\tilde{\hbox{pr}}_2(\hat{W}_i) = Z_i &=& \{ z \in
Z_{i-1} &\kern-10pt/& \kern-10pt\exists\xi \in T_{z}Z_{i-1},\; i_{\xi}\Omega_{L}(z) = 0,\;
\eta_{Z}(z)(\xi) = 1 \},  \end{array}
\]
for all $i \geq 2$. Moreover, one can construct the section $\beta$ of
$Leg_{f}: Z_{f} \longrightarrow \tilde{Z}_{f}$ and the submanifold
$\beta(\tilde{Z}_{f})$ of $Z_{f}$ where a solution of the
Euler-Lagrange equations exists.

The constraint algorithms using equations (\ref{connection1}) and
(\ref{connection2}) and the construction of the corresponding
constraint submanifolds $Z_{i}$ and $\tilde{Z}_{i}$ and of the submanifold
$\beta(\tilde{Z}_{f})$ has been done in \cite{LMM} (see also
\cite{ChLM,barsa}). We remark that, in this case, there exists a
unique solution of the Euler-Lagrange equations on the submanifold
$\beta(\tilde{Z}_{f})$ (for more details, see \cite{LMM}).

\section*{Appendices}

\appendix

\section*{A. Projectable connections.}\label{A}
A connection $\Gamma$ in the fibration
$\pi_{XY} : Y \longrightarrow X$ is given by a horizontal distribution $\hbox{\bf H}$
which is complementary to the vertical one $V\pi_{XY}$, that is
$$
TY = \hbox{\bf H} \oplus V\pi_{XY}.
$$
Associated to the connection there exist a horizontal projector
$\hbox{\bf h} : TY \longrightarrow \hbox{\bf H}$ defined in the obvious manner.

If $(x^\mu, y^i)$ are fibered coordinates, then $\hbox{\bf H}$ is locally
spanned by the local vector fields
$$
(\frac{\partial}{\partial x^\mu})^h =
\frac{\partial}{\partial x^\mu} + \Gamma^i_\mu (x,y) \, \frac{\partial}{\partial y^i} \; ;
$$
$\displaystyle{(\frac{\partial}{\partial x^\mu})^h}$ is called the horizontal lift
of $\displaystyle{\frac{\partial}{\partial x^\mu}}$, and $\Gamma^i_\mu$ are the
Christoffel components of the connection.

Along the paper we repeatedly use the following construction.

Assume that $\pi_{XZ} : Z \longrightarrow X$ and $\pi_{XY} : Y \longrightarrow X$
are two fibrations with the same base manifold $X$, and that
$\Phi : Z \longrightarrow Y$ is a surjective submersion (in other words,
a fibration as well) preserving the fibrations, say,
$\pi_{XY} \circ \Phi = \pi_{XZ}$.

Let $\Gamma$ be a connection in
$\pi_{XZ} : Z \longrightarrow X$ with horizontal projector $\hbox{\bf h}$.

\defin{Definition}{A.1.
$\Gamma$ is said to be projectable if $T\Phi(z) (\hbox{\bf
H}_z)=T\Phi(z') (\hbox{\bf H}_{z'}),$ for all $z,z'\in Z$ in the
same fibre of $\Phi.$ }

If $\Gamma$ is projectable, then we define a connection $\Gamma'$ in the
fibration $\pi_{XY} : Y \longrightarrow X$ as follows:
The horizontal subspace at $y \in Y$ is given by
$$
\bar{\hbox{\bf H}}_y = T\Phi(z) (\hbox{\bf H}_z) \;,
$$
for an arbitrary $z$ in the fibre of $\Phi$ over $y$. It is routine to prove
that $\bar{\hbox{\bf H}}$ defines a horizontal distribution
in the fibration $\pi_{XY} : Y \longrightarrow X$.

We can choose fibered coordinates $(x^\mu, y^i, z^a)$ on $Z$ such that
$(x^\mu, y^i)$ are fibered coordinates on $Y$.
The Christoffel components of $\Gamma$ are obtained by computing the horizontal lift
$$
(\frac{\partial}{\partial x^\mu})^h =
\frac{\partial}{\partial x^\mu} + \Gamma^i_\mu (x,y,z) \, \frac{\partial}{\partial y^i}
+ \Gamma^a_\mu (x,y,z) \, \frac{\partial}{\partial z^a} \;.
$$
A simple computation shows that $\Gamma$ is projectable
if and only if the Christoffel components $\Gamma^i_\mu$ are constant along the fibres
of $\Phi$, say $\Gamma^i_\mu = \Gamma^i_\mu(x,y)$.
In this case, the horizontal lift of $\displaystyle{\frac{\partial}{\partial x^\mu}}$
with respect to $\Gamma'$ is just
$$
(\frac{\partial}{\partial x^\mu})^h =
\frac{\partial}{\partial x^\mu} + \Gamma^i_\mu (x,y) \, \frac{\partial}{\partial y^i} \; .
$$

As an exercise, the reader can easily check that, conversely, given a
connection $\Gamma'$ in the fibration $\pi_{XY} : Y \longrightarrow
X$ and a surjective submersion $\Phi : Z \longrightarrow Y$ preserving
the fibrations, one can construct a connection $\Gamma$ in the fibration
$\pi_{XZ} : Z \longrightarrow X$ which projects onto $\Gamma'$.

\section*{B. Semiholonomic connections.}\label{B}
Let $\pi_{XY} : Y \longrightarrow X$ be a fibration and
$\pi_{XZ} : Z \longrightarrow X$ its 1-jet prolongation, that is, $Z = J^1\pi_{XY}$.
Assume that $X$ is orientable with volume form $\eta$.

\defin{Definition}{B.2.
A connection $\Gamma$ in the fibration
$\pi_{XZ} : Z \longrightarrow X$ is said to be {\rm semiholonomic} if
\begin{equation}\label{sh}
S_\eta(\hbox{\bf h}, \dots, \hbox{\bf h}) = 0,
\end{equation}
where $\hbox{\bf h}$ is the horizontal projector of $\Gamma$.
If (\ref{sh}) holds at a point $z \in Z$, then $\Gamma$ is said to be semiholonomic at $z$.
}

Assume that
$$
\hbox{\bf h}(\frac{\partial}{\partial x^\mu})
= \frac{\partial}{\partial x^\mu} + \Gamma^i_\mu \frac{\partial}{\partial y^i}
+ \Gamma^i_{\mu \nu} \frac{\partial}{\partial z^i_\nu}
$$
in fibered induced coordinates. Then $\Gamma$ is semiholonomic
if and only if we have $\Gamma^i_\mu = z^i_\mu$.

\section*{C. Connections on submanifolds.}\label{C}
The notion of connection in a fibration admits a useful generalization
to submanifolds of the total space.

Let $\pi_{XY} : Y \longrightarrow X$ be a fibration and $P$ a submanifold of $Y$.

\defin{Definition}{C.1.
A connection in $\pi_{XY} : Y \longrightarrow X$ along the submanifold $P$
consists of a family of linear mappings
$$
\hbox{\bf h}_y : T_yY \longrightarrow T_yP
$$
for all $y \in P$, satisfying the following properties
$$
\hbox{\bf h}_y^2 = \hbox{\bf h}_y, \quad
\ker \, \hbox{\bf h}_y = (V\pi_{XY})_y,
$$
for all $y \in P$.
The connection is said to be {\rm differentiable (flat)} if the $n$-dimensional distribution
$\hbox{Im} \,\hbox{\bf h} \subset TP$ is smooth (integrable), where $n = \dim X$.
}

We have the following.

\th{Proposition}{C.2.
Let $\hbox{\bf h}$ a connection in $\pi_{XY} : Y \longrightarrow X$along a submanifold
$P$ of $Y$. Then:
\begin{enumerate}

\item[(1)] $\pi_{XY}(P)$ is an open subset of $X$.

\item[(2)] $(\pi_{XY})_{|_P} : P \longrightarrow \pi_{XY}(P)$ is a fibration.

\item[(3)] The 1-jet prolongation $J^1(\pi_{XY})_{|_P}$ is a submanifold of $Z$.

\item[(4)] There exists an induced true connection $\Gamma_P$ in the fibration
$(\pi_{XY})_{|_P} : P \longrightarrow \pi_{XY}(P)$ with the same horizontal subspaces.

\item[(5)] $\Gamma_P$ is flat if and only if $\hbox{\bf h}$ is flat.

\end{enumerate}

}

\Proof

{\sl (1) \hbox{\rm and} (2)} First of all, we shall prove that $(\pi_{XY})_{|_{P}} : P \longrightarrow X$ is a submersion.

Let $y \in P$ such that $\pi_{XY}(y)=x \in X$. We define a linear mapping
$$
{\cal A}(y) : T_xX \longrightarrow T_yP
$$
as follows:
$$
{\cal A}(y)(U)=\hbox{\bf h}_y(\bar{U}),
$$
where $\bar{U} \in T_yY$ and $T\pi_{XY}(\bar{U}) = U$. The mapping ${\cal A}(y)$
is well-defined since if $\bar{U}'$ is another tangent vector in $T_yY$ satisfying
$T\pi_{XY}(\bar{U}') = U$, then $\bar{U}-\bar{U}' \in (V\pi_{XY})_y$, and therefore
$\hbox{\bf h}_y(\bar{U}')=\hbox{\bf h}_y(\bar{U})$.

In addition, ${\cal A}(y)$ is injective. In fact, if $U \in T_xX$ is such that
${\cal A}(y)(U)=0$, then $\hbox{\bf h}_y(\bar{U}) = 0$, that implies
$\bar{U} \in (V\pi_{XY})_y$, and therefore
$U = T\pi_{XY}(\bar{U}) = 0$.

Finally, ${\cal A}(y)$ is a section of $T\pi_{XY}(y) : T_yP
\longrightarrow T_xX$. Indeed, take $U \in T_xX$; we have ${\cal
A}(y)(T\pi_{XY}({\cal A}(y)(U))) = \hbox{\bf h}_y({\cal A}(y)(U))
= \hbox{\bf h}_y^2 (\bar{U}) = \hbox{\bf h}_y(\bar{U}) = {\cal
A}(y)(U)$. Thus, we have proved that $T\pi_{XY} \circ {\cal A}(y)
= \hbox{Id}_{T_xX}$. This shows that $(\pi_{XY})_{|_P} : P
\longrightarrow X$ is a submersion.

Therefore, $\pi_{XY}(P)$ is an open submanifold of $X$, and
$(\pi_{XY})_{|_P} : P \longrightarrow \pi_{XY}(P)$ is a fibration.

{\sl (3)} is obvious.

{\sl (4)} The induced connection $\Gamma_P$ is defined by restricting
the horizontal subspaces of $\hbox{\bf h}$, that is,
$$
\hbox{\bf h}'_y = (\hbox{\bf h}_y)_{|T_yP}, \quad \hbox{for all} \;
\; y \in P.
$$
Since $\hbox{Im} \, \hbox{\bf h}' = \hbox{Im} \, \hbox{\bf h}$ then {\sl (5)} follows.
\sq

\vskip4pt plus2pt

\end{document}